\documentclass{PoS}
\pdfoutput=1 

\usepackage[table]{xcolor}
\usepackage{colortbl}
\RequirePackage{ifpdf} 
\usepackage{amsmath} 
\usepackage{mathtools}
\usepackage{cases}

\usepackage{pstricks}
\usepackage[final]{pdfpages} 
\usepackage{ifpdf} 
\usepackage{slashed}

\usepackage{color}
\usepackage{xcolor}
\definecolor{urlblue}{rgb}{0.2,0.4,0.7}
\definecolor{citegreen}{rgb}{0,0.6,0.2}
\definecolor{linkred}{rgb}{0.9,0.2,0.1}

\usepackage{graphics}
\usepackage{etoolbox} 
\usepackage{fixmath}
\usepackage{psfrag}

\usepackage{notoccite} 

\usepackage{amsfonts}
\usepackage{autobreak}
\usepackage{marginnote}

\usepackage{tikz}
\usetikzlibrary{positioning,arrows}
\usetikzlibrary{decorations.pathmorphing}
\usetikzlibrary{decorations.markings}
\usetikzlibrary{shapes.geometric}
\tikzset{
    vector/.style={decorate, decoration={snake}, draw},
    provector/.style={decorate, decoration={snake,amplitude=2.5pt}, draw},
    antivector/.style={decorate, decoration={snake,amplitude=-2.5pt}, draw},
    fermion/.style={draw=black, postaction={decorate},decoration={markings,mark=at position .55 with {\arrow[draw=black]{>}}}},
    fermionbar/.style={draw=black, postaction={decorate},
                       decoration={markings,mark=at position .55 with {\arrow[draw=black]{<}}}},
    fermionnoarrow/.style={draw=black},
    gluon/.style={decorate, draw=black,decoration={coil,amplitude=4pt, segment length=5pt}},
    scalar/.style={dashed,draw=black, postaction={decorate},decoration={markings,mark=at position .55 with {\arrow[draw=black]{>}}}},
    scalarbar/.style={dashed,draw=black, postaction={decorate},decoration={markings,mark=at position .55 with {\arrow[draw=black]{<}}}},
    scalarnoarrow/.style={dashed,draw=black},
    electron/.style={draw=black, postaction={decorate},decoration={markings,mark=at position .55 with {\arrow[draw=black]{>}}}},
    bigvector/.style={decorate, decoration={snake,amplitude=4pt}, draw},
}

\title{\vspace*{-5em}
\mbox{}\hfill \mbox{\small\sc IMSc/2019/12/14, MPP-2019-246}\\
\vspace*{6em}
Two-Loop QCD Helicity Amplitudes for Higgs Production Associated with a Vector Boson through Bottom Quark Annihilation}

\ShortTitle{$b{\overline{b}} \rightarrow ZH$}

\author{\speaker{Taushif Ahmed}\thanks{T. Ahmed, L. Chen, P. K. Dhani and V. Ravindran thank the organisers of RADCOR 2019.}
\\
        Max-Planck-Institut f\"ur Physik, Werner-Heisenberg-Institut, 80805 M\"unchen, Germany\\
        E-mail: \email{taushif@mpp.mpg.de}}

\author{A.~H. Ajjath\\
        The Institute of Mathematical Sciences, HBNI, Taramani, Chennai 600113, India\\
        E-mail: \email{ajjathah@imsc.res.in}}
        
\author{Long Chen\\
        Max-Planck-Institut f\"ur Physik, Werner-Heisenberg-Institut, 80805 M\"unchen, Germany\\
        E-mail: \email{longchen@mpp.mpg.de}}

\author{Prasanna K. Dhani\\
        INFN, Sezione di Firenze, I-50019 Sesto Fiorentino, Florence, Italy\\
        E-mail: \email{prasannakumar.dhani@fi.infn.it}}

\author{Pooja Mukherjee\\
        The Institute of Mathematical Sciences, HBNI, Taramani, Chennai 600113, India\\
        E-mail: \email{poojamukherjee@imsc.res.in}}
        
\author{V. Ravindran\\
        The Institute of Mathematical Sciences, HBNI, Taramani, Chennai 600113, India\\
        E-mail: \email{ravindra@imsc.res.in}}

\abstract{
We present the two-loop QCD corrections to the amplitude of the Higgs production associated with a $Z$ boson via the bottom quark-antiquark annihilation channel with a non-vanishing  bottom-quark Yukawa coupling. The computation is performed by projecting the D-dimensional scattering amplitude directly onto a set of Lorentz structures related to the linear polarisation states of the $Z$ boson. We cross-check the finite remainders through a computation based on conventional form factor decomposition. We show that for physical observables, an ultimate D-dimensional form factor decomposition of amplitudes is not necessary which has a huge potential to simplify a multiloop computation. We compute numerically the resulting cross sections under the soft-virtual approximation to NNLO and find it three orders of magnitude smaller than that of the s-channel.
}

\FullConference{14th International Symposium on Radiative Corrections (RADCOR2019)\\ 
9-13 September 2019\\
		Palais des Papes, Avignon, France}

\begin{document}
\allowdisplaybreaks[4]


\def\D{{\cal D}}
\def\DD{\overline{\cal D}}
\def\g{\overline{\cal C}}
\def\gm{\gamma}
\def\M{{\cal M}}
\def\ep{\epsilon}
\def\epm1{\frac{1}{\epsilon}}
\def\epm2{\frac{1}{\epsilon^{2}}}
\def\epm3{\frac{1}{\epsilon^{3}}}
\def\epm4{\frac{1}{\epsilon^{4}}}
\def\unM{\hat{\cal M}}
\def\ashat{\hat{a}_{s}}
\def\asmur{a_{s}^{2}(\mu_{R}^{2})}
\def\sigbar{{{\overline {\sigma}}}\left(a_{s}(\mu_{R}^{2}), L\left(\mu_{R}^{2}, m_{H}^{2}\right)\right)}
\def\sigbarn{{{{\overline \sigma}}_{n}\left(a_{s}(\mu_{R}^{2}) L\left(\mu_{R}^{2}, m_{H}^{2}\right)\right)}}
\def\unas{ \left( \frac{\hat{a}_s}{\mu_0^{\epsilon}} S_{\epsilon} \right) }
\def\rnM{{\cal M}}
\def\bt{\beta}
\def\cD{{\cal D}}
\def\cC{{\cal C}}
\def\ca{\text{\tiny C}_\text{\tiny A}}
\def\cf{\text{\tiny C}_\text{\tiny F}}
\def\ct{{\red []}}
\def\sv{\text{SV}}
\def\murOmu{\left( \frac{\mu_{R}^{2}}{\mu^{2}} \right)}
\def\bb{b{\bar{b}}}
\def\bt0{\beta_{0}}
\def\bt1{\beta_{1}}
\def\bt2{\beta_{2}}
\def\bt3{\beta_{3}}
\def\gm0{\gamma_{0}}
\def\gm1{\gamma_{1}}
\def\gm2{\gamma_{2}}
\def\gm3{\gamma_{3}}
\def\nn{\nonumber}
\def\l{\left}
\def\r{\right}
\def\T{{\cal Z}}    
\def\U{{\cal Y}}

\def\nn{\nonumber\\}
\def\ep{\epsilon}
\def\T{\mathcal{T}}
\def\V{\mathcal{V}}

\newcommand\myeq{\stackrel{\mathclap{\normalfont\mbox{\tiny FR}}}{=}}

\def\qgraf{{\fontfamily{qcr}\selectfont
QGRAF}}
\def\python{{\fontfamily{qcr}\selectfont
PYTHON}}
\def\form{{\fontfamily{qcr}\selectfont
FORM}}
\def\reduze{{\fontfamily{qcr}\selectfont
REDUZE2}}
\def\kira{{\fontfamily{qcr}\selectfont
Kira}}
\def\litered{{\fontfamily{qcr}\selectfont
LiteRed}}
\def\fire{{\fontfamily{qcr}\selectfont
FIRE5}}
\def\air{{\fontfamily{qcr}\selectfont
AIR}}
\def\mint{{\fontfamily{qcr}\selectfont
Mint}}
\def\hepforge{{\fontfamily{qcr}\selectfont
HepForge}}
\def\arXiv{{\fontfamily{qcr}\selectfont
arXiv}}
\def\Python{{\fontfamily{qcr}\selectfont
Python}}
\def\ginac{{\fontfamily{qcr}\selectfont
GiNaC}}
\def\polylogtools{{\fontfamily{qcr}\selectfont
PolyLogTools}}
\def\anci{{\fontfamily{qcr}\selectfont
Finite\_ppbk.m}}
\def\gosam{{\fontfamily{qcr}\selectfont
GoSam}}
\def\fermat{{\fontfamily{qcr}\selectfont
fermat}}
\def\xml{{\fontfamily{qcr}\selectfont
qgraf-xml-drawer}}

\newcommand{\dis}{}
\newcommand{\overbar}[1]{mkern-1.5mu\overline{\mkern-1.5mu#1\mkern-1.5mu}\mkern
1.5mu}
\newcommand{\TODO}[1]{ {\color{red} #1} }

\section{Introduction and Setup}
\label{sec:intro}

\begin{figure}[htbp]
\begin{center}
\includegraphics[scale=0.38]{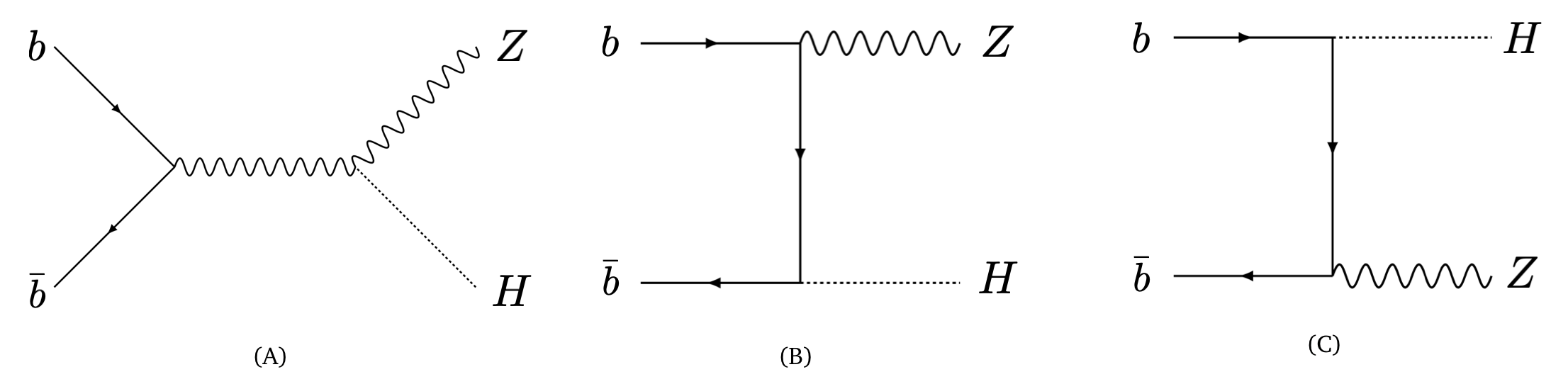}
\caption{Feynman diagrams at leading order}
\label{dia:tree}
\end{center}
\end{figure}

It is thus very desirable to have a precise knowledge about the $VH$ process at hadron colliders, especially to meet the foreseeable precision requirements from future experiments for studies of the 125 GeV Higgs boson as well as potentially non-standard Higgs bosons with ever more details.
In this article, we focus on the  $b$-quark-induced $ZH$ process that involves a non-vanishing Yukawa coupling $\lambda_b$  between the $b$ quark and the Higgs boson:
\begin{align}
\label{eq:process}
    b(p_1) + \bar{b}(p_2) \to Z(q_1) + H(q_2)\,.
\end{align}
$p_i$, $q_i$  represent the momentum of the corresponding particle satisfying on-shell conditions $p_i^2=0,q_1^2=m_z^2,q_2^2=m_h^2$.  $m_z$ and $m_h$ are the mass of the $Z$ and Higgs boson, respectively. At tree level, there are three Feynman diagrams, as shown in figure~\ref{dia:tree}. The diagram (A) gives an $s$-channel contribution with the same structure as that of the Drell-Yan production, which has been studied extensively in QCD to order $\mathcal{O}(\alpha_s^3)$ in refs.~\cite{Ahmed:2014cla,Li:2014bfa,Catani:2014uta,Kumar:2014uwa}. At the tree level there are three contributing diagrams for the $b$-quark-induced $ZH$ process\footnote{This holds in the physical unitary gauge where unphysical degree-of-freedoms in electroweak ghosts and would-be-Goldstone bosons decouple from the spectrum.}, as shown in figure~\ref{dia:tree}. In this article, we focus on the diagrams (B) and (C) that involve $\lambda_b$ and compute the two-loop corrections in massless QCD with $n_f=5$ flavour scheme under dimensional regularisation.

\noindent
The kinematic invariants are defined as 
\begin{align}
\label{eq:Mandelstam}
    s\equiv (p_1+p_2)^2\,,~~ t\equiv (p_1-q_1)^2\, ~~\text{and}~~ u\equiv (p_2-q_1)^2
\end{align}
satisfying $s+t+u=q_1^2+q_2^2=m_z^2+m_h^2$. 
%
We keep a non-zero Yukawa coupling between Higgs boson and $b$ quark, otherwise the mass of the $b$-quark is set to be zero. 
For the axial coupling between $Z$-boson and massless quarks, we define $\gamma_5$ in dimensional regularisation~\cite{tHooft:1972tcz,Bollini:1972ui} (DR) as~\cite{tHooft:1972tcz,Breitenlohner:1977hr}
\begin{align}
\label{eq:gamma5}
	\gamma_5=\frac{i}{4!}\varepsilon_{\mu\nu\rho\sigma}\gamma^{\mu}\gamma^{\nu}\gamma^{\rho}\gamma^{\sigma}\,.
\end{align}
As a consequence of this definition, we need to symmetrise the axial current~\cite{Akyeampong:1973xi,Larin:1991tj} and upon doing so, we obtain~\cite{Akyeampong:1973xi,Larin:1991tj}
\begin{align}
\label{eq:gamma5-axial}
	\gamma_{\mu}\gamma_5 = \frac{i}{6} \varepsilon_{\mu\nu\rho\sigma} \gamma^{\nu} \gamma^{\rho} \gamma^{\sigma}\,,
\end{align}
which is used in D-dimensions for our calculation. Moreover, all the Lorentz indices of the Levi-Civita symbols are considered in D- dimensions~\cite{Zijlstra:1992kj}. In the next section, we introduce the projector method that is adopted in this article.

\section{Prescription of Projectors}
\label{sec:projectors}
In this work, the polarised amplitudes of the process \eqref{eq:process} are obtained by projecting the D-dimensional amplitude directly onto a minimal set of D-dimensional projection operators following the approach proposed in~\cite{Chen:2019wyb}. In order to cross-check our calculation, we also compute the polarised amplitudes by following the conventional method of decomposing the amplitude into independent Lorentz structures and then applying appropriate projectors in D-dimensions to extract the corresponding form factors. In this section, we present the recipe for both the methods in brief.

\subsection{Projectors for Linearly Polarised $b \bar{b} Z H$ Amplitudes}
\label{sec:projectors_LP}

The polarisation projectors as introduced in~\cite{Chen:2019wyb} are based on the momentum basis representations of external state vectors (for both bosons and fermions, massless or massive), and all their open Lorentz indices are by definition taken to be D-dimensional to facilitate a uniform projection with just one dimensionality D=$g_{~\mu}^{\mu}$. This allows us to find a Lorentz covariant representation of the tensor products of external particles' state vectors solely in terms of external momentum and algebraic constants, such as the metric tensor, spinor matrices and Levi-Civita symbol. As all the open Lorentz indices are considered to be D-dimensional, no dimensional splitting ever comes into the picture for loop momenta and/or Lorentz indices of inner vector fields.

The amplitude of \eqref{eq:process} can be schematically parameterised as
\begin{align} 
\label{eq:bbZHamplitude}
    \mathcal{M} &= \bar{v}(p_2) \, {\Gamma}^{\mu} \, u(p_1) \, \varepsilon^{*}_{\mu}(q_1) \nonumber\\
    &= \bar{v}(p_2) \, {\Gamma}^{\mu}_{vec} \, u(p_1) \, \varepsilon^{*}_{\mu}(q_1) 
    \,+\,\bar{v}(p_2) \, {\Gamma}^{\mu}_{axi} \, u(p_1) \, \varepsilon^{*}_{\mu}(q_1)\nonumber\\
    &\equiv {\cal M}_{vec}+{\cal M}_{axi}\,.
\end{align}
The symbol ${\Gamma}^{\mu} \equiv {\Gamma}^{\mu}_{vec} + {\Gamma}^{\mu}_{axi}$ stands for a matrix in the spinor space with one open Lorentz index $\mu$ which may be carried by either the elementary Dirac matrix $\gamma^{\mu}$ or one of the external momentum. 
It consists of contributions from the vector and axial coupling of the $Z$ boson, denoted  by 
${\Gamma}^{\mu}_{vec}$ and ${\Gamma}^{\mu}_{axi}$, respectively.
Due to the presence of Yukawa coupling $\lambda_b$ vertex on the external $b$ quark line, all non-vanishing amplitudes involve two external massless spinors, $u(p_1)$ and $v(p_2)$, with opposite \textit{chirality}. As a consequence, the power of spinor matrices in ${\Gamma}^{\mu}$ sandwiched between the two external massless $b$ quark spinors must be even in order to have a non-vanishing matrix element between such a pair of spinors.

Following the ref.\cite{Chen:2019wyb}, the projectors are obtained as
\begin{align} 
\label{eq:LPprojectors_primitive}
    \bar{u}(p_1) \, {\mathbf{N}}_i \, v(p_2) \, \varepsilon^{\mu}_j \, , 
    \quad \text{for $i = s, p$ and $j = T,Y,L$} 
\end{align}
where the ${\mathbf{N}}_s = \mathbf{1}$, ${\mathbf{N}}_p = \gamma_5$, and $\varepsilon^{\mu}_j$ with $j = T,Y,L$ represent the three linear polarisation eigenstates of $Z$-boson identified in the center-of-mass frame of the collision. We choose $\varepsilon^{\mu}_T$ to be the transversal polarisation within the scattering plane determined by the three linearly independent external momenta, $p_1, p_2, q_1$. The other transversal polarisation $\varepsilon^{\mu}_Y$ 
is orthogonal to $p_1, p_2,$ and $ q_1$, and is constructed with the help of Levi-Civita symbol. 
The third physical polarisation state of the $Z$ boson, its longitudinal polarisation which is denoted by the vector $\varepsilon^{\mu}_L$, has its spatial part aligned with its own momentum $q_1$.     To get the form of the \eqref{eq:LPprojectors_primitive}, we first write down a Lorentz covariant ansatz for the $\varepsilon^{\mu}_j$ and then solve the orthogonality and normalisation conditions of linear polarisation state vectors for the linear decomposition coefficients. Once we establish a definite Lorentz covariant decomposition form in 4 dimensions solely in terms of external momenta and kinematic invariants, this form is declared as the \textit{definition} of the corresponding polarisation state vector in D dimensions. By substituting the obtained results of the polarisation tensor and the definition of $\gamma_5$ according to \eqref{eq:gamma5}, and then performing the simplification using 4-dimensional Lorentz and spinor algebra, we \textit{choose} a set of 6 projectors, $ \mathcal{P}_i^{\mu}$  to be used in D-dimensions. The explicit form of the projectors can be found in~\cite{Ahmed:2019udm}. We emphasise that since by construction the momentum basis representations of polarisation state vectors fulfil all the defining physical constraints, hence the index contraction between the projectors  and the amplitude \eqref{eq:bbZHamplitude} is simply done with the space-time metric tensor $g_{\mu\nu}$ instead of physical polarisation sum. From the linearly polarised amplitudes obtained by applying the aforementioned projectors, we can get the helicity (with respect to massless $b$-quarks) amplitudes as 
\begin{align} 
\label{eq:HLamps}
    {\mathcal{M}}^{[\pm\pm T]} &= \mathcal{N}_{\pm\pm T}^{-1} \, 
    \Big( \mathcal{P}^{\mu}_1 \,\mp\,  \mathcal{P}^{\mu}_4 \Big)\, \bar{v}(p_2) \, {\Gamma}_{\mu} \, u(p_1) \, , \nonumber\\
    \mathcal{M}^{[\pm\pm Y]} &= \mathcal{N}_{\pm\pm Y}^{-1} \, 
    \Big( \mathcal{P}^{\mu}_2 \,\mp\,  \mathcal{P}^{\mu}_5 \Big)\, \bar{v}(p_2) \, {\Gamma}_{\mu} \, u(p_1) \, , \nonumber\\
    \mathcal{M}^{[\pm\pm L]} &= \mathcal{N}_{\pm\pm L}^{-1} \, 
    \Big( \mathcal{P}^{\mu}_3 \,\mp\,  \mathcal{P}^{\mu}_6 \Big)\, \bar{v}(p_2) \, {\Gamma}_{\mu} \, u(p_1) \, , 
\end{align}
where the superscript, e.g. ++$T$, denotes the respective polarisations of the $b$ quark, anti-$b$ quark and the $Z$ boson. The normalisation factors are explicitly given in ref. \cite{Ahmed:2019udm}. Each helicity amplitude can be decomposed into vector and axial parts. Moreover, these can be expanded in powers of strong coupling constant $a_s=\alpha_s/4\pi$. In this article, we compute these quantities to second order in $a_s$.

\subsection{Projectors for Conventional Form Factors}
\label{sec:projectors_FF}
In order to cross check the results obtained using the linearly polarised projectors, we perform the calculation using conventional form factor decomposition which we now turn into. The vector part of the amplitude can be decomposed in terms of 4 linearly independent Lorentz structures as
\begin{align} 
\label{eq:FFD_vec}
\bar{v}(p_2) \, {\Gamma}^{\mu}_{vec} \, u(p_1) &= 
    F_{1,vec} \, \bar{v}(p_2) \, u(p_1) \, p_1^{\mu} 
    \,+\, F_{2,vec} \, \bar{v}(p_2) \, u(p_1) \, p_2^{\mu} \nonumber\\
    &\,+\, F_{3,vec} \, \bar{v}(p_2) \, u(p_1) \, q_1^{\mu}
    \,+\, F_{4,vec} \, \bar{v}(p_2) \,\gamma^{\mu} \slashed{q}_1 u(p_1) \,,
\end{align}
which is followed by the constraints of the presence of even powers of elementary Dirac matrices, combined with the P-even constraint from the vector coupling. The coefficients $F_{i,vec}$ are called vector form factors.  The aforementioned basis of Lorentz structures is complete in D-dimensions even without demanding the transversality of the $Z$ boson's physical polarisation states, since we prefer to use the simple metric tensor $g_{\mu\nu}$ 
instead of the physical polarisation sum rule for index contraction in projections. By computing the inverse of the Gram matrix of the aforementioned basis, we obtain the explicit form of the projectors for the above form factor decomposition which can be found from~\cite{Ahmed:2019udm}. By applying the derived projectors onto the Feynman diagrams, we obtain the vector part of the amplitudes.

For the form factor decomposition of the axial part of the amplitude, as a result of using the D-dimensional definition of $\gamma_5$ which forces us to sacrifice the anti-commuting property, a linearly independent and complete Lorentz structures in 4-dimensions does not necessarily hold true in D-dimensions. This is because while constructing the basis structures in 4-dimensions, the anti-commuting property of $\gamma_5$ is routinely employed which fails in D-dimensions. To begin with, we consider the linearly independent and complete basis in 4-dimensions as
\begin{align} 
\label{eq:FFD_axi_4}
\Big\{
\bar{v}(p_2) \,\gamma_5\, u(p_1) \, p_1^{\mu} \,,\, 
\bar{v}(p_2) \,\gamma_5\, u(p_1) \, p_2^{\mu} \,,\, 
\bar{v}(p_2) \,\gamma_5\, u(p_1) \, q_1^{\mu} \,,\, 
\bar{v}(p_2) \,\gamma^{\mu} \gamma_5\, \slashed{q}_1 u(p_1)
\Big\} \,,
\end{align}
which is constructed under the condition of even powers in elementary Dirac matrices combined with the P-odd constraint from the axial coupling. By substituting the D-dimensional definition of $\gamma_5$, we calculate the D-dimensional projectors through computing the inverse of the Gram matrix whose explicit forms can be found from~\cite{Ahmed:2019udm}. 

Now we address the question whether the vector and axial parts of the amplitudes, ${\cal M}_{vec}$ and ${\cal M}_{axi}$, live in a space spanned by the aforementioned Lorentz structures depicted through \eqref{eq:FFD_vec} and \eqref{eq:FFD_axi_4} in D-dimensions. In order to check it, we calculate the tree level amplitude in D-dimensions directly from Feynman diagrams, where we make use of the D-dimensional definition of non-anticommuting $\gamma_5$, and append the resulting Lorentz structures to the original set \eqref{eq:FFD_vec} and \eqref{eq:FFD_axi_4}. Now we find that the rank of the Gram matrix of the vector part remains 4 whereas for the axial it gets increased to 5. This clearly implies that in D-dimensions the tree-level vector amplitude ${\cal M}_{vec}$ indeed lives in a space spanned by the four Lorentz structures depicted through  \eqref{eq:FFD_vec}, whereas the tree-level axial amplitude ${\cal M}_{axi}$ is not fully spanned by the corresponding structures in \eqref{eq:FFD_axi_4}. Following the same argument, even if we manage to get a complete basis for the tree level amplitude in D-dimensions, it is not guaranteed to retain the completeness property at loop level with the same set of structures. So, in general, for axial amplitude it is not easy to construct a complete set of basis in D-dimensions such that it remains so even at higher loop order. Hence, we can not write down an exact all order form factor decomposition of the axial part, ${\cal M}_{axi}$, which could be analogous to the vector counterpart, ${\cal M}_{vec}$ in \eqref{eq:FFD_vec}. 

So, the question arises if we start our computation with an incomplete basis of Lorentz structures in D-dimensions, as depicted through \eqref{eq:FFD_axi_4}, would it always lead to correct results, because for sure the so reconstructed amplitude is not algebraically identical to the original defining form given by Feynman diagrams. On top of this, there is also no clear statement in the literature whether one can always set the dimension variable D = 4 in the expressions of the projectors, and still expect with confidence that the amplitude reconstructed in this way yields correct results of the physical observables. Keeping the full D-dependence on the axial projectors generally leads to a cumbersome set of projectors. 
With this motivation, we start by decomposing the axial part of the amplitude to linearly independent and complete structures in 4-dimensions but incomplete in D-dimensions, as presented in  \eqref{eq:FFD_axi_4}, and construct the projectors. In the expressions of the projectors, we set D=4 and  \textit{define} the ``axial form factors'' as
\begin{align} 
\label{eq:FFD_axi_def}
F_{1,axi} &\equiv {\mathbb{P}}^{[\text{4}],\,\mu}_{1,axi} \, \bar{v}(p_2) \, {\Gamma}^{\nu}_{axi} \, u(p_1)\,g_{\mu \nu}\,\nonumber\\
F_{2,axi} &\equiv {\mathbb{P}}^{[\text{4}],\,\mu}_{2,axi} \, \bar{v}(p_2) \, {\Gamma}^{\nu}_{axi} \, u(p_1)\,g_{\mu \nu}\,\nonumber\\
F_{3,axi} &\equiv {\mathbb{P}}^{[\text{4}],\,\mu}_{3,axi} \, \bar{v}(p_2) \, {\Gamma}^{\nu}_{axi} \, u(p_1)\,g_{\mu \nu}\,\nonumber\\
F_{4,axi} &\equiv {\mathbb{P}}^{[\text{4}],\,\mu}_{4,axi} \, \bar{v}(p_2) \, {\Gamma}^{\nu}_{axi} \, u(p_1)\,g_{\mu \nu}\,
\end{align}
where the $[\text{4}]$ signifies  D = 4 in the original set of projectors ${\mathbb{P}}^{[\text{4}],\,\mu}_{i,axi}$. Note that, however, when applying these projectors to $\mathcal{M}_{axi}$, all the Lorentz and spinor algebra including the contraction of pair of Levi-Civita symbols must be performed in D-dimensions. Consequently, we build up an intermediate axial amplitude $\tilde{\mathcal{M}}_{axi}^{\mu}$ which is defined as 
\begin{align} 
\label{eq:FFD_axi_amp}
\tilde{\mathcal{M}}_{axi}^{\mu} &\equiv 
    F_{1,axi} \, \bar{v}(p_2) \,\gamma_5\, u(p_1) \, p_1^{\mu} 
    \,+\, F_{2,axi} \, \bar{v}(p_2) \,\gamma_5\, u(p_1) \, p_2^{\mu} \nonumber\\
    &\,+\, F_{3,axi} \, \bar{v}(p_2) \,\gamma_5\, u(p_1) \, q_1^{\mu}
    \,+\, F_{4,axi} \, \bar{v}(p_2) \,\gamma^{\mu} \, \gamma_5\, \slashed{q}_1 u(p_1) \,.
\end{align}
This quantity $\tilde{\mathcal{M}}_{axi}^{\mu}$ is not algebraically identical to the original amplitude $\mathcal{M}_{axi}^{\mu} = \bar{v}(p_2) \, {\Gamma}^{\mu}_{axi} \, u(p_1)$ in D dimensions, while we expect that the two should eventually lead to the same properly defined finite remainder in 4-dimensions and consequently, lead to correct physical observables.

Indeed, by comparing with the results for polarised amplitudes computed in this article using physical projectors defined in section~\ref{sec:projectors_LP}, not relying on any explicit Lorentz tensor decomposition of the original Feynman amplitude, we have verified that eventually the same finite remainders are obtained for the axial part of the amplitude hence we confirm a positive answer to the question raised above.
In conclusion, for all practical purposes where we confine ourselves in the computation of physical observables, we are allowed to decompose the amplitudes in D-dimensions, be it vector or axial, into a set of Lorentz structures which are linearly independent and complete in 4-dimensions and then construct the corresponding projectors (i.e. a set that is essentially identical to the respective expressions that would be used in a four-dimensional form factor decomposition). Applying these 4-dimensional projectors onto the Feynman diagrams and performing everything in D-dimensions lead to correct finite remainder. We expect this crucial observation to hold true even beyond the amplitude considered in this article.


\section{UV Renormalisation}
\label{sec:uv}

The bare amplitudes beyond leading order (LO) are ultraviolet (UV) divergent whose rectification requires not only the renormalisation of strong and Yukawa coupling constants, but also the counterterms arising from the axial currents as a result of using non-anticommuting $\gamma_5$. The renormalisations of $a_s$ and $\lambda_b$ are performed in $\overline{\rm MS}$ scheme~\cite{Tarasov:1980au}. The flavour non-singlet axial current
\begin{align}
\label{eq:ns-axial-current}
    J^{ns}_{\mu,A}(x) = \bar\psi(x) \gamma_{\mu}\gamma_5  I_{3} \psi(x)=\frac{i}{6} \epsilon_{\mu\nu\rho\sigma} \bar\psi(x) \gamma^{\nu}\gamma^{\rho}\gamma^{\sigma} I_3 \psi(x) \, ,
\end{align}
where $I_3$ denotes the third component of the (weak) isospin operator, is renormalised in $\overline{\rm MS}$  through the multiplication of an overall operator renormalisation constant along with a finite renormalisation~\cite{Trueman:1979en,Larin:1991tj}
\begin{align}
\label{eq:renorm-ns-current}
    J^{ns}_{\mu,A}(x) = Z^{ns}_{5,A} Z^{ns}_A {\hat J}^{ns}_{\mu,A}(x)\,.
\end{align}
The hat ( ${\hat{}}$ ) represents the bare quantity. The expressions for the aforementioned renormalisation constants can be found from~\cite{Ahmed:2019udm}. The UV renormalised non-singlet current does not exhibit the Ward identity. However, this is not an anomalous property, this happens due to the presence of a non-zero Yukawa coupling. The failure of holding the Ward identity for non-singlet current is dictated by classical expectation.

The flavour-singlet (or anomalous) diagrams, characterised by featuring a triangle fermion loop with axial $Z$ coupling, start to appear at two-loop. 
Due to Furry's theorem, only the axial part of these diagrams survives. The singlet current is defined as
\begin{align}
\label{eq:axial-current}
    J^{s}_{\mu,A}(x) = \bar\psi(x) \gamma_{\mu}\gamma_5  \psi(x)=\frac{i}{6} \epsilon_{\mu\nu\rho\sigma} \bar\psi(x) \gamma^{\nu}\gamma^{\rho}\gamma^{\sigma} \psi(x)\,.
\end{align}
Similar to the non-singlet current, this is also renormalised through~\cite{Kodaira:1979pa,Collins:1984xc,Larin:1993tq}:
\begin{align}
\label{eq:renorm-s-current}
    J^{s}_{\mu,A}(x) = Z^{s}_{5,A} Z^{s}_A {\hat J}^{s}_{\mu,A}(x)\,,
\end{align}
where the renormalisation constants can be found explicitly in ref. \cite{Ahmed:2019udm}. In contrast to the flavour non-singlet axial current, the singlet  current does not satisfy the standard Ward identity even in the massless quark limit. In other words, this exhibits anomalous properties, known as axial or Adler-Bell-Jackiw (ABJ) anomaly~\cite{Adler:1969gk,Bell:1969ts}. The operator relation for the ABJ anomaly of massless axial current reads~\cite{Adler:1969er}
\begin{align}
\label{eq:ABJ}
    \left(\partial^{\mu} J^{s}_{\mu,A}\right)_R=a_s\frac{1}{2}\left(G\tilde G\right)_R \, ,
\end{align}
where $G\tilde G \equiv \epsilon_{\mu\nu\rho\sigma}G^a_{\mu\nu}G^a_{\rho\sigma}$ and $G^a_{\mu\nu}$ is the gluonic field strength tensor. The subscript $R$ represents that these composite local operators need to be properly renormalised.  We have verified this operator level relation explicitly within our computational setup for the singlet axial amplitude which provides an indirect check to our treatment of these anomalous diagrams.


\section{Computation of the Amplitudes}
\label{sec:computation}

The technicalities of the computation of bare amplitudes closely follow the steps used in the recent calculation of two-point form factors with two-operator insertion with different virtualities in maximally supersymmetric Yang-Mills theory~\cite{Ahmed:2019upm}. Feynman diagrams are generated symbolically using \qgraf~\cite{Nogueira:1991ex}. At two-loops, there are 153 flavour non-singlet and 6 flavour singlet diagrams are present. Using in-house routines based on \form~\cite{Vermaseren:2000nd} are used to perform the SU(N) colour, spinor and Lorentz algebras in D-dimensions. Upon applying the projectors onto the Feynman diagrams, we end up with thousands of scalar Feynman integrals which are reduced to a much smaller subset of integrals, called master integrals (MI) with the help of integration-by-parts (IBP)~\cite{Tkachov:1981wb,Chetyrkin:1981qh} identities through  \kira~\cite{Maierhoefer:2017hyi,Maierhofer:2018gpa}.  In a parallel setup, using an extension of \gosam~\cite{Cullen:2011ac,Cullen:2014yla,Jones:2016bci}, we obtain the matrix elements. The bare amplitudes in terms of MI are found to agree in both the setup. We use the optimised solutions of the MI computed in the article~\cite{Gehrmann:2015ora} and get the result of the amplitudes. The MI are also computed in refs. \cite{Henn:2014lfa,Caola:2014lpa,Papadopoulos:2014hla}. The analytic results of the UV renormalised polarised amplitudes to two-loops are attached as ancillary files with the \arXiv~submission of ref. \cite{Ahmed:2019udm}.

\section{IR Factorisation and RS Independent Finite Remainders}
\label{sec:ir}

The UV renormalised amplitudes are found to exhibit the expected behaviour~\cite{Catani:1998bh,Becher:2009cu}  of the soft and collinear (IR) divergences in conventional dimensional regularisation (CDR) which serves as the most stringent check on our computation. 

We have checked explicitly that there is an exact agreement of the vector parts of the six bare amplitudes (with full D-dependence) between the direct projection calculation and the one with a detour of conventional D-dimensional form factor decomposition. Consequently, the 4-dimensional finite remainders are identical in both the methods.

However, for the axial part, the projectors for projecting our ``axial form factors'' defined in \eqref{eq:FFD_axi_def} are a bit un-conventional due to two points as already mentioned 
(1) the basis set \eqref{eq:FFD_axi_4} is known to be linearly incomplete in D dimensions for $\mathcal{M}_{axi}$ in question;
(2) all explicit D appearing in the corresponding set of projectors are set manually to be 4. This makes the axial form factor decomposition not qualified as being called D-dimensional faithful representation. Still we defined an intermediate axial amplitude $\tilde{\mathcal{M}}_{axi}$ in \eqref{eq:FFD_axi_amp} from ``axial form factors''. It would be very interesting to see whether the correct 4-dimensional finite remainders ${\cal M}^{[j]}_{axi,\mathrm{fin}}$ could still be obtained in a computation based on such an acrobatic version of axial form factor decomposition. We find that the 4-dimensional finite remainders in both the methods do agree which implies it is not necessary to construct and use the ``ultimate'' D-dimensional axial decomposition basis, as long as one is only concerned with physical quantities. Using the analytical results of the polarised amplitudes, in the next section, we compute the inclusive production cross-section at NNLO QCD at threshold.

\section{Cross-Section to NNLO in Soft-Virtual Approximation}
\label{sec:sv}

The hadronic cross section of the $ZH$ production that results from the $t,u$-channels of $b$ quark initiated partonic sub-processes is given by
\begin{align}
\label{eq:Xsection}
    \sigma^{ZH}_{tu} = \sum_{a=b,\overline b}\int dx_1 f_{a} (x_1,\mu_F) \int dx_2 {f}_{\bar a}(x_2,\mu_F)  
\sigma^{ZH}_{a \bar a,tu}(x_1,x_2,m_z,m_h,\mu_F)\,.
\end{align}
Here, $x_{1,2}$ are fractions of momenta of incoming hadrons carried away by bottom quarks $b$ and $\bar b$, and ${f}_{a}$ is the parton distribution function (PDF) normalised at the factorisation scale $\mu_F$.  The mass factorised partonic cross section is given by ${\sigma}^{ZH}_{b\bar b,tu}$. In absence of the results of all the real emission diagrams, we expand the cross-section around threshold limit $z \equiv Q^2/s=1$ where $Q^2=(q_1+q_2)^2$ is the invariant mass square of the final state $ZH$ and obtain the leading contribution to the cross-section which is called soft-virtual (SV) approximation. Owing to the universality of the soft contributions, we compute the required counterterm by following the methodology described in~\cite{Ravindran:2005vv,Ravindran:2006cg,H:2018hqz} and consequently, obtain the SV cross-section at NNLO QCD where only the contributions of the kind $\delta(1-z)$ and $D_j(z)\equiv\left(\frac{\log^j(1-z)}{(1-z)} \right)_+$ are retained. At the 13 TeV LHC, we compute the numerical contributions arising from the SV cross-section and we find that these $t,u$-channels give a three orders of magnitude smaller contribution compared to that of the $s$-channel one. This contribution can be enhanced in theories like MSSM where the Yukawa coupling can be enhanced.

\section{Conclusions}
\label{sec:con}

Through this work we provide the analytic results of the two-loop massless QCD corrections to the $b$-quark-induced $ZH$ process involving non-vanishing $b$-quark Yukawa coupling $\lambda_b$, which is a necessary ingredient of the complete  $\mathcal{O}(\alpha_s^2)$ QCD corrections to this process in the five-flavour scheme. The computation was performed in two different methods. In the one based on momentum basis representation of the linear polarisation vector~\cite{Chen:2019wyb}, we constructed the linearly polarised projectors and applied them directly onto the Feynman diagrams to obtain the polarised amplitudes. The finite remainders are cross-checked against the respective results obtained through the conventional method of form factor decomposition. We confirm that as long as we confine ourselves to the computation of physical observables it is not necessary to construct and use the ``ultimate'' D-dimensional form factor decomposition basis of the amplitudes. An acrobatic version of form factor decomposition is shown to be sufficient in this computation done with the axial current regularised in dimensional regularisation, which we think should also hold in other cases. This conclusion has the potential to simplify multiloop calculation, particularly those involving axial coupling. As expected, we find that the $t,u$-channels give a three orders of magnitude smaller contribution compared to that of the $s$-channel one towards the SV cross-section at NNLO.

\bibliography{bbzh} 
\bibliographystyle{JHEP}
\end{document}